# Three Dimensional Liquid Gated Graphene Field Effect Strain Sensor

Shideh K. Ameri, Pramod K. Singh, and Sameer R. Sonkusale

*Abstract*— Realizing flexible strain sensor with high sensitivity and tunable gauge factor is a challenge. To meet this challenge, we report an ionic liquid gated three-dimensional graphene field effect strain sensor. The charge carrier concentration in this 3D graphene is modulated by applied electric field through an all-around self-assembled electrical double layer capacitance formed at the interface of graphene with ionic liquid. Strain causes folding and unfolding of microscopic wrinkles and formation of cracks in the graphene network altering transistor behavior. Mechanical deformation of graphene also alters its bandgap providing inherent strain sensitivity. Use of 3D network results in robust operation since there exists multiple paths for the charge carriers to flow between source and drain terminals. Interestingly, changing the applied bias allows one to tune the gauge factor of this graphene transistor based strain sensor. The current-voltage characteristics of the sensor were measured for different tensile strain values of 0.5% to 35%. Our results show that the sensor maintains its field effect characteristics over a large strain range; moreover it enables up to 68% tunability in the strain gauge factor. We also report cyclic measurements with varied magnitude and frequency showing repeatability and robustness as a highly sensitive strain sensor.

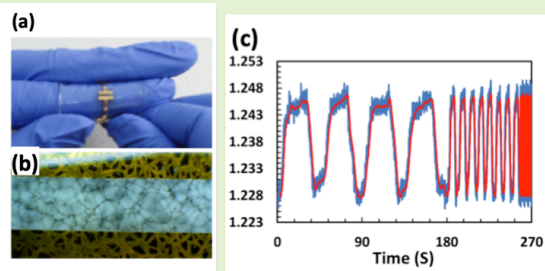

## I. INTRODUCTION

Strain sensors are used in wide range of applications such as monitoring of human motion, structural health, orthopaedic, and tissue implants, and broadly for studying material fatigue in both natural and synthetic systems.[1-6] Conventional strain sensors based on semiconductors and metals are low in cost and have high sensitivity, but they are usually unidirectional and quite bulky in size. In comparison, using nano-materials for strain sensing have the advantages of scalability, high sensitivity and multi-directionality. Various nano-materials have been investigated for strain sensing applications such as zinc oxide nanowires, zinc-stannate (ZnSnO.) nanowires, carbon nanotubes (CNT), and graphene layers, flakes and nano-sheets. Different values of gauge factor (GF) which defines the sensitivity of the strain sensor ranging between 0.06 to 10. have been reported in the literature.[7-12] Among these nanomaterials, graphene has impressive mechanical stability, high carrier mobility and electrical conductivity. Furthermore, it is transparent, biocompatible and non-hazardous which make it an attractive material for strain sensing especially for biological applications. Graphene film based strain sensors have similar or greater gauge factor compared to the silver nano-wire based sensors (GF of 2-14) or conventional metal based sensors (GF of 2-5).[12-14] However, gauge factor of graphene is much lower than ZnO nanowires (GF of 116), CNT (GF of 0.82-176) and ZnSnO3 nanowires (GF of 3750) based strain sensors.[8, 9, 15-17] Graphene sheet when processed in to different forms and physical structures such as micro-ribbons, micro and nano-flakes, graphene ripples, woven graphene and graphene foams possess different electrical properties than planar graphene film. The GF of graphene film based strain sensors has been reported to be between 0.11 to 6.6, rippled graphene 2, micro-ribbon 9.49, nano-graphene film 300 and graphene woven fabric between 10. to 10.[10, 15, 18, 19] The scalability of the gauge factor is highly desirable as specific gauge factors are needed for certain applications. The high gauge factor sensors are suitable for low strain applications, while low gauge factor sensors are applicable for high strain applications. However, there is a trade-off between elasticity and gauge factor in graphene-based strain sensors since the mechanism of the sensing in these types of sensors is based on the formation of cracks during stretching. For example, woven graphene gauge factor of 10. for the tensile strain range of up to 10%, while the super elastic rippled graphene shows a considerably higher dynamic range of tensile strain up to 30% but significantly lower GF of 2.[10, 18] The higher elasticity leads to a smaller gauge factor and vice versa. The use of passive multilayer free-standing graphene foam for strain sensing has been reported. It has been shown that multilayer graphene foam/PDMS composite is highly stretchable and it can be strained up to 90%.[20] However, due to high electrical conductivity and stretchability of multilayer graphene foam it does not show high GF at low strains.

This work was supported in part by the National Science Foundation (NSF) under 1931978 and 1240443."

S.K.Ameri was with Tufts University, Medford, MA 2155, USA. She is now with Department of Electrical and Computer Engineering, Queen's University, Kingston, ON K7L 3N6, Canada. (e-mail: shideh.ameri@queensu.ca).

P.K. Singh was with Tufts University, Medford, MA 02155, USA. He is now with with ³Aelius Semiconductors Pvt. Ltd., Singapore (e-mail: pramodsingh.mt@gmail.com).

S. Sonkusale is with the Department of Electrical and Computer Engineering Tufts University, Medford MA 02155 USA (e-mail: sameer@ece.tufts.edu). S.Sonkusale is the corresponding author.

## II. EXPERIMENTAL METHODS

To achieve high sensitivity at low strains, while maintain good sensitivity at higher strains, we made a field effect strain sensor consisting of few-layers of graphene (1-3 layers) in the form of 3D networked foam [21]. We show such a network capable of exhibiting tunable GF for low and high strains. The 3D interconnected graphene networks provide multiple parallel pathways for electrons and holes to conduct current and therefore ensuring sensor's fault-tolerance in the presence of cracks and discontinuities due to excessive applied strain. The 3D graphene sensor is a field effect active strain sensor where the carrier concentration and therefore the sensitivity and GF is tunable by applying electric field to the graphene through ionic liquid that surrounds the graphene. Our results show that the conductivity of the channel decreases with increasing strain, resulting in decrease of overall drain-source current in graphene transistor based sensor. Graphene sensor maintains its field effect characteristics and ambipolar behavior up to 35% of applied strain. The sensor shows the gauge factors from 1.89 to 16.6 depending on the level of tensile strain, which is higher than graphene film, rippled graphene and micro-ribbon based strain sensors; and the GF values can be tuned up to 68% by using electric field. In the following sections, strain sensing mechanism, fabrication and performance of this sensor will be discussed.

### A. Sensing Mechanism

A large group of flexible strain sensors are passive strain gauges. The electrical conductivity of strain gauges changes by reversible mechanical deformation (when they are stretched below their permanent deformation point) that can be translated to the magnitude of applied strain. Gauge factor (GF) is a quantitative factor, which is used to define the sensitivity of a strain sensor to the applied strain, and it is defined as;

$$GF = \frac{\Delta R}{\varepsilon R_0} \qquad (1)$$

where, $\Delta R$ is change in the electrical resistance due to strain, $R_0$ is initial resistance and $\varepsilon$ is applied strain which is defined as $\Delta L / L_0$. The parameters, $\Delta L$ and $L_0$ are change in the length due to strain and initial length, respectively.

The presented field effect strain sensor is an active strain gauge sensor whose gauge factor and responsivity can be tuned. Unlike metals where the high electron charge density influences its electrical conductivity, in graphene both electrons and holes act as electrical carriers albeit at lower concentrations that influence its material conductivity. Applying electric field alters its Fermi level and consequently the number of free carriers, making graphene a field-effect device. When graphene is used as strain gauge, the magnitude of the applied electric field directly influences its electrical conductivity affecting its gauge factor. The mechanical deformation and strain of the graphene foam results in unfolding of its microscopic wrinkles altering the effective geometric length of the device, whereas further strain of graphene also causes elongation of carbon-carbon bonds in its 2D lattice altering its bandgap. In extreme strain, cracks and discontinuities may also result. Thus the magnitude of the applied strain to the graphene based sensor can therefore be translated to the change in the electrical conductivity of the transistor based sensor under different bias conditions. The interesting synergy of mechanical dependence of graphene's bandgap, geometric folding/unfolding of the wrinkles or cracking, and bias dependent field effect transistor behaviour, make the 3D graphene transistor as an excellent candidate for an active strain sensor with tunable GF.

### B. Fabrication

The steps in the fabrication of the three-dimensional graphene based strain sensor are shown in Fig. 1. In the first step graphene grown on 3D copper foam is left in copper etchant to etch away the underlying copper (Fig. 1a). Then the 3D graphene network is rinsed with DI water and transferred to polydimethylsiloxane (PDMS) substrate (Fig. 1b). Next, two metal contacts (Cr/Au) are formed and encapsulated by SU8 photoresist (Fig. 1c & d). Finally ionic liquid (1-Butyl-3-methylimidazolium hexafluorophosphate, BMIM-PF6, 98%) was added on graphene network (Fig. 1e) for liquid gating. The image of fabricated device is shown in the inset of Fig. 1(e). Figure 2a shows the photo of the flexible and stretchable sensor. The optical image of the graphene network immersed in the ionic liquid is shown in Fig. 2b. The strain sensor was characterized by applying various level of tensile strain up to 45% using customized set-up as shown in Fig. 2(c). The sensor was stretched using two metal clips connected to a moving stage with micrometer movement precision while the electric field was applied to graphene using an Ag/AgCl electrode in ionic liquid serving as gate. At the interface of graphene and ionic liquid, a double layer capacitance forms which serves as an all around gate capacitance in field effect transistor. The electrical transfer characteristics of the device at different level of applied strain were measured using

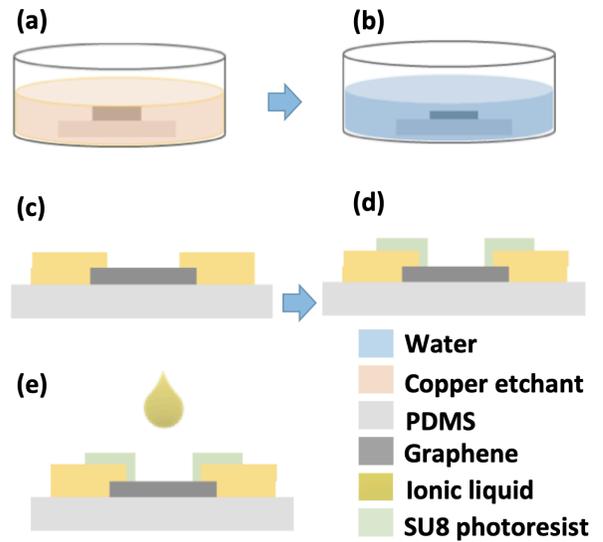

Fig. 1. The fabrication process of 3D graphene foam based strain sensor. a) Etching underlying copper in graphene network on copper, using copper etchant, b) rinsing the graphene network by DI water, c) forming Cr/Au contacts, d) encapsulating the contacts using SU8 photoresist, e) immersing graphene in ionic liquid for liquid gating. Ag/AgCl electrode (not shown) is immersed in ionic liquid to apply gate potential.

semiconductor parameter analyzer (HP Agilent 4156A).

### III. RESULTS AND DISCUSSION

Figure 2d shows the change in the conductivity of the channel by applying 5% to 45% strain. Initially unfolding of wrinkles results in change in effective geometric length providing strain sensitivity. At higher strains, dependence of graphene bandgap on mechanical deformation in addition to formation of cracks and discontinuities results in increasing electrical resistance. As a result, the current flowing through the graphene decreases when voltage is applied between to metal contacts under increasing strain. The graphene bipolar

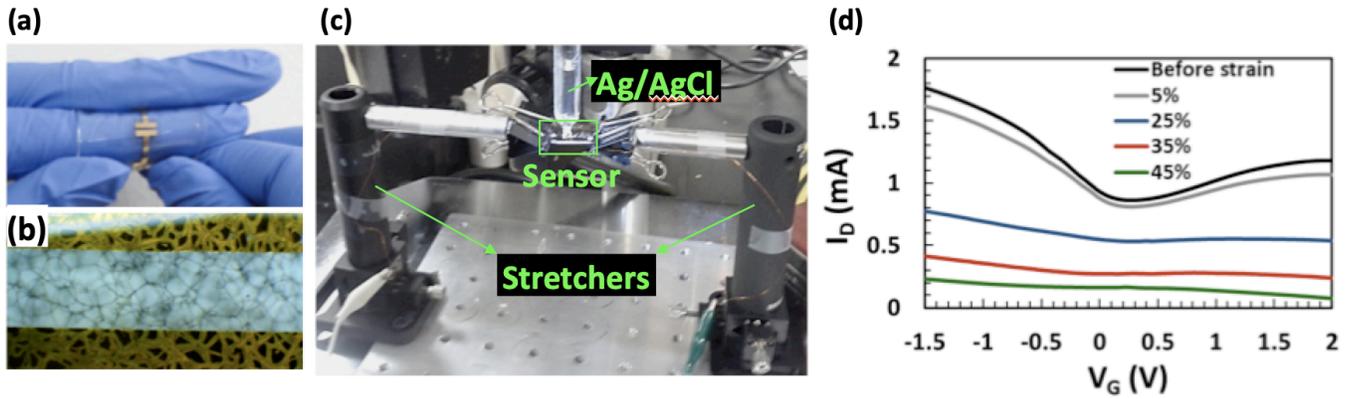

**(a)** **(c)** Ag/AgCl Sensor Stretchers **(d)**

Fig. 2. a) Mechanically flexible graphene based field effect strain sensor, b) the optical image of network of graphene ribbons immersed in ionic liquid (blue part graphene in ionic liquid, the yellow parts Cr/Au contacts), c) the photo of the customized set-up for measuring the current-voltage characteristics of the strain sensor, d) transfer characteristics of the field effect strain sensor at different level of tensile strain (0%, 5%, 25%, 35% and 45%).

behavior was clearly observed when the fermi level of graphene changed by applying voltage to the ionic liquid surrounding the graphene. Therefore, the sensor gauge factor can be adjusted by applying electric field to graphene through double layer capacitance.

The 3D network of the graphene ribbons provides different possible pathways for the electrons and holes to flow in presence of potential cracks and discontinuity in some ribbons resulting in the stable and robust operation of the sensor at high percentage of strain. The cyclic tests were performed using Instron 3369, interfaced with PC-based data acquisition and analysis software, LABVIEW, to investigate the stability and repeatability of the sensor response to the strain. Figure 3 shows the cyclic measurement results under saw-tooth wave strain at different percentages of applied strain from 0.5% to 12%. Results show that the response and recovery of the sensor are stable and repeatable under various level of strain, suggesting the potential application of the sensor for sensing both low and high level of tensile strains.

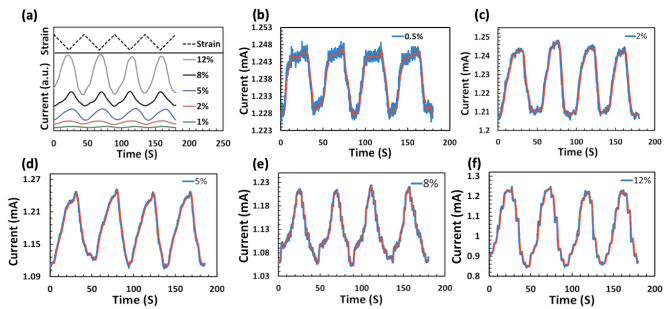

Fig. 3. Sensor's cyclic strain test after applying different level of tensile strain from 0.5% to 12%.

Figure 4(a) shows the gauge factor of the device versus different percentages of tensile strain at -1.5 V applied voltage to ionic liquid surrounding the graphene ribbons network. The gauge factors measured as 1.89 at 5%, and 16.6 at 45% of tensile strain. It has been shown that the network of interconnected graphene ribbons can be modeled as a combination of the series and parallel resistors and formation of cracks in the graphene network due to strain results in exponential increase in the overall electrical resistance of the sensor, but there is an order of magnitude difference between experimental and calculated results[Error! Bookmark not defined.]. This raises the probability of associated tunnelling effect at higher values of tensile strain through nanometer gaps generated due to nano-cracks in the graphene caused by strain. This will be basis of future investigation.

Figure 4b shows the tunability of the gauge factor by applying electric field to the graphene ribbon network immersed in ionic liquid. Results shows that the tunability of up to 68% can be achieved at 5% tensile strain, 48% tunability can be achieved at 25% of strain and 35.5% at 35% of applied tensile strain.

The results of cyclic measurements at different frequencies, 0.02 Hz, 0.2 Hz and 0.5 Hz, at 0.5 % of the strain is presented in Fig. 4c. This result shows the similar response to the strain at different cycling frequencies and indicates that the performance of sensor is repeatable and stable over different frequencies of alternative strain.

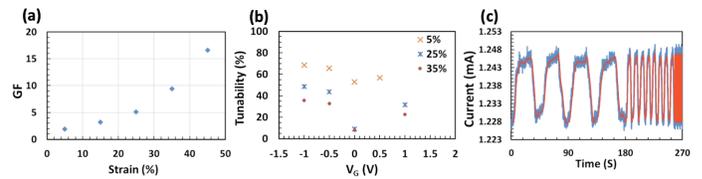

Fig. 4. a) The plot of strain sensor's gauge factor versus applied tensile strain, b) the plot of GF tunability versus applied voltage to the immersed graphene network in ionic liquid at different level of applied tensile strain, c) cyclic measurement of the strain sensor at different speeds for 0.5% of tensile strain.

## IV. CONCLUSION

A 3D graphene based field effect strain sensor with tunable gauge factor (GF) up to 68% is reported. The sensor show robust operation over a wide dynamic range of tensile strain from 0.5% to 35% during which the 3D networked graphene transistor shows a strain-dependent ambipolar I-V characteristic behavior. The sensor is robust since the 3D network of graphene ribbons ensures connectivity even in the presence of discontinuities or cracks that may appear at high strain levels. The measured gauge factor is 1.89 to 16.6 at different levels of applied tensile strain. The device shows stable and repeatable functionality even under repeated cyclic test with tensile frequencies from 0.02 Hz to 0.5 Hz. The mechanism of sensing in this device is a combination of geometric unfolding of wrinkles, mechanical dependence of graphene's bandgap, and the formation of defects in graphene under increased strain level that modulates graphene's conductance. There is also a possibility of associated tunneling effect of electrons through the gaps formed by cracks at high tensile strain. Inherent transistor based amplification and tunability of gauge factor makes this 3D graphene-based strain sensor useful in many applications as wearables in monitoring human motion [9, 22, 23] or as implants to monitor fatigue in biomedical implants. [24]

## ACKNOWLEDGMENT

Authors acknowledge the financial support of the National Science Foundation and Office of Naval Research for this work. Authors would also like to thank the Tufts Micro and Nano Fabrication facility.